\input harvmac

\skip0=\baselineskip

\def\encadremath#1{\vbox{\hrule\hbox{\vrule\kern8pt\vbox{\kern8pt
\hbox{$\displaystyle #1$}\kern8pt} \kern8pt\vrule}\hrule}}

\def\p{\partial}

% approximately less than

% approximately greater than

 \lref\gt{
  G.~W.~Gibbons and P.~K.~Townsend,
  ``Black holes and Calogero models,''
  Phys.\ Lett.\  B {\bf 454}, 187 (1999)
  [arXiv:hep-th/9812034].
  %%CITATION = PHLTA,B454,187;%%
}
%\SimonsNM
\lref\ssty{ A.~Simons, A.~Strominger, D.~M.~Thompson and X.~Yin,
  ``Supersymmetric branes in AdS(2) x S**2 x CY(3),''
  Phys.\ Rev.\ D {\bf 71}, 066008 (2005)
  [arXiv:hep-th/0406121].
  %%CITATION = HEP-TH 0406121;%%
  %%Cited 7 times in SPIRES-HEP
}

%\GauntlettNW
\lref\GauntlettNW{
  J.~P.~Gauntlett, J.~B.~Gutowski, C.~M.~Hull, S.~Pakis and H.~S.~Reall,
  ``All supersymmetric solutions of minimal supergravity in five dimensions,''
  Class.\ Quant.\ Grav.\  {\bf 20}, 4587 (2003)
  [arXiv:hep-th/0209114].
  %%CITATION = HEP-TH 0209114;%%
}
\lref\gvtwo{
  R.~Gopakumar and C.~Vafa,
  ``M-theory and topological strings. II,''
  arXiv:hep-th/9812127.
  %%CITATION = HEP-TH 9812127;%%
  %%Cited 97 times in SPIRES-HEP
}
%\GopakumarII
\lref\gvone{
 R.~Gopakumar and C.~Vafa,
``M-theory and topological strings. I,''
arXiv:hep-th/9809187.
%%CITATION = HEP-TH 9809187;%%
}

%\VafaGR
\lref\VafaGR{ C.~Vafa, ``Black holes and Calabi-Yau threefolds,''
Adv.\ Theor.\ Math.\ Phys.\  {\bf 2}, 207 (1998)
[arXiv:hep-th/9711067].
%%CITATION = HEP-TH 9711067;%%
}

\lref\osv{
  H.~Ooguri, A.~Strominger and C.~Vafa,
  ``Black hole attractors and the topological string,''
  Phys.\ Rev.\ D {\bf 70}, 106007 (2004)
  [arXiv:hep-th/0405146].
  %%CITATION = HEP-TH 0405146;%%
  %%Cited 89 times in SPIRES-HEP
}

\lref\msw{  J.~M.~Maldacena, A.~Strominger and E.~Witten,
  ``Black hole entropy in M-theory,''
  JHEP {\bf 9712}, 002 (1997)
  [arXiv:hep-th/9711053].
  %%CITATION = HEP-TH 9711053;%%
  %%Cited 96 times in SPIRES-HEP
}

\lref\ssty{ A.~Simons, A.~Strominger, D.~M.~Thompson and X.~Yin,
``Supersymmetric branes in AdS(2) x S**2 x CY(3),''
arXiv:hep-th/0406121.
%%CITATION = HEP-TH 0406121;%%
}

\lref\bu{
  R.~Dijkgraaf, R.~Gopakumar, H.~Ooguri and C.~Vafa,
  ``Baby universes in string theory,''
  arXiv:hep-th/0504221.
  %%CITATION = HEP-TH 0504221;%%
  %%Cited 23 times in SPIRES-HEP
}

\lref\mina{
  M.~Aganagic, H.~Ooguri, N.~Saulina and C.~Vafa,
  ``Black holes, q-deformed 2d Yang-Mills, and non-perturbative topological
  strings,''
  Nucl.\ Phys.\ B {\bf 715}, 304 (2005)
  [arXiv:hep-th/0411280].
  %%CITATION = HEP-TH 0411280;%%
  %%Cited 34 times in SPIRES-HEP
}

%\BatesVX
\lref\BatesVX{
  B.~Bates and F.~Denef,
  ``Exact solutions for supersymmetric stationary black hole composites,''
  arXiv:hep-th/0304094.
  %%CITATION = HEP-TH/0304094;%%
}

\lref\dnf{
  F.~Denef,
  ``Supergravity flows and D-brane stability,''
  JHEP {\bf 0008}, 050 (2000)
  [arXiv:hep-th/0005049].
  %%CITATION = HEP-TH 0005049;%%
}

\lref\dm{
 F.~Denef and G.~W.~Moore,
  ``Split states, entropy enigmas, holes and halos,''
  arXiv:hep-th/0702146.
  %%CITATION = HEP-TH/0702146;%%
} \lref\cv{C.~Vafa,
  ``Two dimensional Yang-Mills, black holes and topological strings,''
  arXiv:hep-th/0406058.}

\lref\gravmul{
  A.~Fujii, R.~Kemmoku and S.~Mizoguchi,
  ``D = 5 simple supergravity on AdS(3) x S(2) and N = 4 superconformal  field
  theory,''
  Nucl.\ Phys.\ B {\bf 574}, 691 (2000)
  [arXiv:hep-th/9811147].
  %%CITATION = HEP-TH 9811147;%%
  }
  \lref\gsyads{
  D.~Gaiotto, A.~Strominger and X.~Yin,
  ``From AdS(3)/CFT(2) to black holes / topological strings,''
  arXiv:hep-th/0602046.
  %%CITATION = HEP-TH/0602046;%%
  }

  \lref\iceland{
  R.~Britto-Pacumio, J.~Michelson, A.~Strominger and A.~Volovich,
  `Lectures on superconformal quantum mechanics and multi-black hole  moduli
  spaces,'' proceedings of the Akuyreri, Iceland summer school on
  string theory August, 1999,
  arXiv:hep-th/9911066.
  %%CITATION = HEP-TH/9911066;%%
}

\lref\bsv{
  R.~Britto-Pacumio, A.~Strominger and A.~Volovich,
  ``Two-black-hole bound states,''
  JHEP {\bf 0103}, 050 (2001)
  [arXiv:hep-th/0004017].
  %%CITATION = JHEPA,0103,050;%%
}
\lref\sm{
  S.~D.~Mathur,
  ``The fuzzball proposal for black holes: An elementary review,''
  Fortsch.\ Phys.\  {\bf 53}, 793 (2005)
  [arXiv:hep-th/0502050].}
  \lref\jmas{
  J.~M.~Maldacena and A.~Strominger,
  ``AdS(3) black holes and a stringy exclusion principle,''
  JHEP {\bf 9812}, 005 (1998)
  [arXiv:hep-th/9804085].
  %%CITATION = JHEPA,9812,005;%%
}
%\MateosQS
\lref\MateosQS{
  D.~Mateos and P.~K.~Townsend,
  ``Supertubes,''
  Phys.\ Rev.\ Lett.\  {\bf 87}, 011602 (2001)
  [arXiv:hep-th/0103030].
  %%CITATION = HEP-TH 0103030;%%
}
\lref\gsyscqm{
  D.~Gaiotto, A.~Strominger and X.~Yin,
  ``Superconformal black hole quantum mechanics,''
  JHEP {\bf 0511}, 017 (2005)
  [arXiv:hep-th/0412322].
  %%CITATION = JHEPA,0511,017;%%
}

%\draft

\Title{\vbox{\baselineskip12pt\hbox{} }} { Black Hole
Deconstruction} \centerline{Frederik Denef\footnote{$^*$}{Institute
for Theoretical Physics, K.U. Leuven, Celestijnenlaan 200D, B-3001
Leuven, Belgium}, Davide Gaiotto\footnote{$^\dagger$}{Center for the
Fundamental  Laws of Nature, Jefferson Physical Laboratory, Harvard
University, Cambridge, MA  USA} }\centerline{ Andrew
Strominger$^\dagger$, Dieter Van den Bleeken$^*$ and Xi Yin$^\dagger
$}
\smallskip
\centerline{} \vskip .6in \centerline{\bf Abstract} {A $D4$-$D0$
black hole can be deconstructed into a bound state of  $D0$ branes
with a $D6$-$\bar D6$ pair containing worldvolume fluxes.  The exact
spacetime solution is known and resembles a $D0$ accretion disk
surrounding a $D6$-$\bar D6$ core. We find a scaling limit in which
the disk and core drop inside an $AdS_2$ throat. Crossing this
$AdS_2$ throat and the $D0$ accretion disk into the core, we find a
second scaling region describing the  $D6$-$\bar D6$ pair. It is
shown that the M-theory lift of this region is  $AdS_3\times S^2$.
Surprisingly, time translations in the far asymptotic region reduce
to global, rather than Poincar\'e,  time translations in this core
$AdS_3$.  We further find that the quantum mechanical ground state
degeneracy reproduces the Bekenstein-Hawking entropy-area law.
 } \vskip .3in

%\draft
%\smallskip
\Date{}

%\listtoc

%\writetoc

\newsec{Introduction}

A BPS black hole can be deconstructed in a variety of ways into
zero-entropy, minimally charged bits - often wrapped branes. It is
natural to suppose that black hole entropy is the ground state
degeneracy of the quantum mechanics  on the  moduli space of such
deconstructions.   This quantum mechanics is expected to have a
superconformal symmetry and provide the $CFT_1$ dual of the
near-horizon $AdS_2$ geometry. The program to understand  black
holes in this way has never fully succeeded, although the hope that
it may some day do so remains very much alive. Some reviews can be
found in \refs{\iceland, \sm} and some recent progress in\gsyscqm.

Prior attempts appear to have been missing a crucial ingredient
brought to light in \dnf.  Up to now such analyses have mainly
concerned like-charged (or mutually supersymmetric) black holes,
which enjoy a no-force condition and can lie at any relative
location. This leads to various volume divergences and singularities
in the moduli space quantum mechanics. In \dnf\ it was discovered
that in the generic situation, when the black hole charge vectors
point in different directions,  there is a rich, but surprisingly
tractable moduli space of solutions in which all constituents are
bound.  In this paper we will see that in some cases there is a
scaling limit with a superconformal symmetry.  This moduli space
quantum mechanics is qualitatively different from the parallel
charged case, and looks promising for an understanding of black hole
microscopics.

In this paper we will consider the deconstruction of  a single
center charge $(P, q_0)$ $D4-D0$ black hole into a multicenter
system consisting of a $D6$ brane with flux $F=P/2$ on the
worldvolume, a $\bar D6$ with flux $F=-P/2$, and $\tilde
q_0=q_0+{P^3 \over 24}$ $D0$ branes, where $P \gg 1$. For simplicity
we take the dimension of the D4 and D2 charge spaces to be one, but
generalization is straightforward. This system appeared in \dm\ in
the context of a derivation of the OSV conjecture, in which mainly
the case $\tilde{q}_0 \ll P^3$ was important. In this regime, a
single centered black hole of charge $(P,q_0)$ does not exist, and
the centers stay well separated from each other. However when
$\tilde q_0$ becomes greater than $P^3/12$, a phase transition
occurs: the BPS configuration moduli space develops a branch in
which the centers can approach each other arbitrarily close in
coordinate distance, leading to a capped off $AdS_2$ throat of
arbitrary depth. Such configurations are asymptotically
indistinguishable from, and perhaps should be physically identified
with, those of a single centered black hole. In this scaling limit
there is a superconformal symmetry acting on the moduli space
quantum mechanics, which one hopes is related to or identified with
the $CFT_1$ dual of the the $AdS_2$ throat.

In section 2 we review the full asymptotically flat $D0$-$D6$-$\bar
D6$ classical solution in terms of harmonic functions on $R^3$. In
section 3 we write down the scaling  solution, which is effectively
obtained by simply dropping the constants in the harmonic functions
(this is consistent only when certain charge constraints are
satisfied). In the far region, the scaling solution looks like
$AdS_2\times S^2$ and lifts to $AdS_3\times S^2$ in $M$-theory. In
the interior there are pointlike $D0$ brane singularities which are
free to move as long as they stay on an ``accretion disk'' which
lies on the plane equidistant to the $D6$ and $\bar D6$. In the
central ``near region'', far inside the accretion disk, one finds a
geometry whose $M$-lift is again $AdS_3\times S^2$, but now with
{\it global} $AdS_3$. The $D6$ and $\bar D6$ are the north and south
poles of the $S^2$.\foot{This lift of $D6-\bar D6$ to $AdS_3\times
S^2$ also relates the derivations of OSV given in \dm, where the two
factors of $Z_{top}$ come from a $D6$ and a $\bar D6$, to that of
\gsyads\ where they come from the north and south poles of an
$S^2$.} In section 4 we analyze the (super)symmetries of the
solution, and find a nontrivial relation between the generators of
the near and far $AdS_3$s. In particular the twisted global time
generator $\bar L_0^{\rm near}-J^{3 \rm near} $ of the near region
is the Poincare time generator of the far $AdS_3$ as well as the
Hamiltonian of the full asymptotically flat geometry.  This
surprising fact  may provide a derivation of the conjecture
\refs{\gt,\bsv} that the black hole entropy is the degeneracy of
quantum eigenstates with respect to the global time generator of the
near horizon $AdS_2$. In section 5 we analyze supersymmetric branes
which wrap the $S^2$ ``horizon'' and correspond to D0 branes blown
up via the Myers effect.
 In section 6,
following earlier work \gsyscqm, we show that such wrapped branes
have a huge degeneracy, and can account for the Bekenstein-Hawking
entropy. This is a positive indication that the superconformal
quantum mechanics of the deconstruction given here may give a good
picture of the $D4$-$D0$ quantum black hole.

\newsec{The full classical solution}

In this section we present the asymptotically flat four-dimensional
classical solution corresponding to $\tilde q_0$ $D0$-branes and a
$D6$-$\bar D6$ pair with fluxes inducing $D4$-$D2$-$D0$ charges $({P
\over 2}, \pm {P^2 \over 8},-{P^3 \over 48})$, as well as its
five-dimensinal M-theory lift. This is a special case of a more
general class of solutions described in \BatesVX. For notational
simplicity we consider the case of a single vector multiplet (two
$U(1)$s) but the results are easily generalized.

The solution is constructed from the four harmonic functions on
$R^3$
\eqn\rzft{\eqalign{ H^0 &= h^0+{1 \over |x-x_{D6}|}-{1
\over|x-x_{\bar D6}|} ,\cr H^1 &= h^1+{P \over 2}{1 \over
|x-x_{D6}|}+{P \over 2}{1 \over|x-x_{\bar D6}|},\cr H_1 &=h_1 -{P^2
\over 8}{1 \over |x-x_{D6}|}+{P^2 \over 8}{1 \over|x-x_{\bar
D6}|},\cr H_0 &=h_0 -{P^3 \over 48}{1 \over |x-x_{D6}|}-{P^3 \over
48}{1 \over|x-x_{\bar D6}|} + \sum_{i=1}^{\tilde q_0} {1
\over|x-x_{D0}^i|},}}
in a self-evident notation. An integrability condition (the
existence of $\omega$ below) restricts the brane positions to obey
the constraints
\eqn\xzk{ h^0+{1 \over|x_{D6}-x_{D0}^i|}= {1 \over|x_{\bar
D6}-x_{D0}^i|},} \eqn\xzw{(h_0-{P\over 2} h_1-{P^2\over8}
h^1+{P^3\over48} h^0)+\sum_{i=1}^{\tilde q_0} {1
\over|x_{D6}-x_{D0}^i|}={P^3 \over 6 |x_{\bar D6}-x_{D6}|}.}
From the harmonic functions \rzft\ we form the algebraic
combinations
\eqn\qexp{Q = H_1+{(H^1)^2\over 2H^0},}\eqn\jexp{J = -{H_0\over 2}+
{(H^1)^3\over 6(H^0)^2}+{H^1 H_1\over 2 H^0}, }
and
\eqn\llk{S=2\pi \sqrt{ {2\over 9}H^0Q^3 - (H^0J)^2}.}
In terms of these quantities the  complex Calabi-Yau modulus is
\eqn\dsza{z={\pi H^1+i {\p S \over \p H_1} \over \pi H^0+i{\p S
\over \p H_0}}.}
The 4D metric is
\eqn\dsi{ds_4^2=-{\pi \over S}(dt+\omega )^2+{S \over \pi} d \vec
x^2\,,}
$\omega$ here is the 1-form that satisfies
\eqn\omegaa{ *^3d\omega = H_0 dH^0+H_1 dH^1 - H^0 dH_0-H^1 dH_1. }
The gauge fields strengths are
\eqn\sio{dA^\Lambda=d\bigl( S^{-1} {\p S \over \p
H_\Lambda}(dt+\omega) \bigr)+*^3dH^\Lambda ,  }
for $\Lambda=0,1$.

The M-theory lift is given by
\eqn\metric{ ds^2 = -{9\over 8}Q^{-2} (dt+\omega +
2J(dx^5+\omega^0))^2 +Q (H^0 d\vec x^2 +
(H^0)^{-1}(dx^5+\omega^0)^2),}
where $\omega^0$ is determined by
 $d\omega^0=*^3 dH^0$.

\newsec{The scaling solutions}

In this section we describe the ``far-horizon'' and ``near-horizon''
scaling regions, both of which M-lift to (quotients of) $AdS_3\times
S^2$.

We are looking for ``scaling solutions'' obtained by rescaling all
coordinates $x \to \lambda \tilde{x}$ and taking $\lambda \to 0$
while keeping $\tilde{x}$ finite. This effectively amounts to
dropping the constant terms $h$ in the harmonic functions \rzft. The
charges are then sitting at different locations far down a very long
throat.

%The harmonic functions for the scaling solution are:
%%
%\eqn\rft{\eqalign{ H^0 &= {1 \over |x-x_{D6}|}-{1 \over|x-x_{\bar
%D6}|} ,\cr H^1 &= {P \over 2}{1 \over |x-x_{D6}|}+{P \over 2}{1
%\over|x-x_{\bar D6}|},\cr H_1 &= -{P^2 \over 8}{1 \over
%|x-x_{D6}|}+{P^2 \over 8}{1 \over|x-x_{\bar D6}|},\cr H_0 &= -{P^3
%\over 48}{1 \over |x-x_{D6}|}-{P^3 \over 48}{1 \over|x-x_{\bar D6}|}
%+ \sum_{i=1}^{\tilde q_0} {1 \over|x-x_{D0}^i|}.}}
%%
In this scaling limit, the integrability conditions \xzk\ at the
$D0$ location force the $D0$-branes to be equidistant from the $D6$
and the $\bar D6$. Hence all $\tilde q_0$ of them lie in a plane,
say $x^3=0$. resembling an accretion disk or ring surrounding the
$D6$-$\bar D6$ pair. The integrability condition \xzw\ at the $D6$
location fixes the distance $R_6$ between the $D6$ and this plane.
Hence the moduli space of solutions is the $2\tilde q_0$ positions
of the $D0$s on $R^2$.

Since the distance between any D0 and the D6 is at least $R_6$,
\xzw\ implies $\tilde q_0 \geq P^3/12$, so we only have scaling
solutions when this constraint on the charges is met. We are in
particular interested in the regime
\eqn\faz{\tilde q_0 >>P^3/12.}
In this case generically the $D6$ and $\bar D6$ will be much closer
to the plane than to the $D0$s. For example if the $D0$s all lie on
a ring of radius $R_0$ the distance $R_6$ between the $D6$ and the
plane satisfies
\eqn\rfdt{{P^3 \over 12} {1 \over R_6} = {\tilde q_0 \over
\sqrt{R_0^2 + R_6^2}} }
and $R_0 >> R_6$.While the full geometry is pretty complicated, it
simplifies in the far region $|x|>>R_0$ and near region $|x|<<R_0$.

\subsec{Far region}

Far outside the $D0$ ring,  the fine structure of the pointlike
branes disappears down the throat and the geometry following from
\rzft\ with $h=0$ is simply the near horizon geometry of a $D4$-$D0$
black hole. The M-theory lift is
\eqn\rft{{(24 \tilde q_0 -P^3) \over 8P} dx_5^2 + {3 r \over P} dt
dx_5 + {P^2 \over 2 r^2} dr^2 +{P^2 \over 2} (d\eta^2+\sin^2 \eta
d\psi^2), }
where we have traded $\vec x$ for $r,\eta,\psi$. The periodic
identification $x^5\sim x^5+{4\pi}$ makes this an $SL(2,R)_L$
quotient of $AdS_3\times S^2$ .  On the boundary $r\to \infty$, the
conformal metric is simply
\eqn\dsrt{ds_2^2=dtdx^5,}
where $SL(2,R)_L$ ($SL(2,R)_R$) acts in the standard fashion on the
left (right) null coordinate $x^5$ ($t$). In particular we see that
the Hamiltonian $\p_t$ of the original asymptotically flat region
generates null translations of the unbroken $SL(2,R)_R$.

$AdS_3$ quotients of this type were analyzed in \jmas. While $t$ is
a standard null coordinate on the Poincare diamond of the boundary
of $AdS_3$, $x^5$ in fact turns out to be a Rindler coordinate.  The
$SL(2,R)_L$-invariant $AdS_3$ vacuum then gives, after $x^5$
identifications, a thermal state for the left movers of the boundary
CFT. The entropy of this thermal state agrees with that of the
$D4$-$D0$ black hole.

\subsec{Near region}

Now we consider the near region, far inside the accretion disk. It
is convenient to work in prolate spheroidal coordinates
$\rho,\eta,\phi$ (plus $x^5$ and $t$):
\eqn\prolate{ \eqalign{ & |x-x_{D6}| = R_6(\cosh \rho + \cos\eta),
\cr & |x-x_{\bar D6}| = R_6(\cosh\rho - \cos\eta). }
 }
Lines of constant $\rho$ are ellipsoids with foci at the $D6$ and
$\bar D6$ locations, while $\eta$ is an angle around the ellipsoid.
In this coordinate system, the functions appearing in the metric are
 \eqn\qjex{ Q = -{P^2\over 4R_6\cos\eta},}
 \eqn\iiy{ J = {P^3\over 24R_6}{\cosh\rho\over
 \cos^2\eta}-{1\over 2} \sum_{i=1}^{\tilde q_0} {1\over |x-x^i_{D0}|}. }
The near region is characterized by
\eqn\dti{|x|<<|x^i_{D0}|\sim R_0 .}
This allows us to replace the summation in the rhs of \iiy\ by a constant. This constant is fixed by the the integrability condition \xzw.
One then has
\eqn\iiy{ J = {P^3\over 24R_6}\bigl({\cosh\rho\over
 \cos^2\eta}-1\bigr).}
The last term in $H_0$ reduces to the constant  ${P^3 \over 12
R_6}$, and $\omega$ becomes
\eqn\omge{ \omega = {P^3\over 3R_6}
{\sinh^2({\rho\over 2})\sin^2\eta d\phi\over (\cosh^2\rho -
\cos^2\eta)} . }
After  a linear change of coordinates
\eqn\srft{\eqalign{t &= {P^3 \over 3 R_6} \tau,\cr x_5 &= 2 (\tau +
\theta), \cr \phi &= \psi+\tau-\theta\,,}}
the metric \metric\ takes the  global $AdS_3 \times S^2$ form
\eqn\raft{{P^2 \over 2}(d \eta^2 + \sin^2 \eta d\psi^2) + {2P^2
}(-\cosh^2 ({\rho \over 2}) d\tau^2 + \sinh^2 {\rho \over 2}
d\theta^2 + {1 \over 4} d\rho^2)}
with no identifications except the standard $2\pi$ identifications
of $\psi$ and $\theta$. This near region metric has an enhanced
$SL(2,R)_L$$\times SU(1,1|2)$ superisometry group.

\newsec{Supersymmetry}

Calabi-Yau compactification of Type IIA or M theory preserves eight
supersymmetries. Of these, the general multicenter solution
preserves four which we denote
\eqn\trj{~^\pm Q^\pm,}
where the left (right) superscript denotes the charges under shifts
of $x^5$ ($\phi$).  Commutators of these preserved charges give the
Hamiltonian ($i\p_t$)
 \eqn\razx{\{^AQ^\alpha,^BQ^\beta
\}=\epsilon^{AB}\epsilon^{\alpha\beta}H+{\rm central~~terms},}
where $\alpha, A=\pm$. The supercharges commute with $H$:
\eqn\wsz{[^\pm Q^\pm,H]=0.}
In the near and far $AdS$ regions \rft\ and \raft, the four
supercharges are locally enhanced to eight $SU(1,1|2)$
superconformal symmetries. We wish to understand which four of the
eight supersymmetries in each of the near and far regions are
globally preserved.

\subsec{Near region}

Let us denote the eight supercharges of the near region in a
standard convention
\eqn\rtyp{~^{L_0}G^{J^3}_{\bar L_0}=^{\pm}G^{\pm}_{\pm \half}.}
The $G$s have eigenvalues $\pm \half$ under the action of
\eqn\szc{\eqalign{\bar L^{\rm near}_0&=\half(\p_\tau-\p_\theta),\cr
                     L^{\rm near}_0&=\half(\p_\tau+\p_\theta),\cr
                      J_{\rm near}^3&=\p_\psi  ,}}
Using the coordinate transformation  \srft\ we find
\eqn\wju{{P^3 \over 3 R_6}H=2(\bar L^{\rm near}_0-J_{\rm near}^3).}
The four linear combinations of the eight $^\pm G^\pm_{\pm \half}$
which correspond to $~^\pm Q^\pm$ can then be deduced by demanding
that they commute with $\bar L^{\rm near}_0-J_{\rm near}^3$ as in
\wsz. One finds
\eqn\prs{~^\pm Q^+=^\pm G^+_{\half}, ~~~~^\pm Q^-=^\pm
G^-_{-\half}.}

\subsec{Far Region}

In the far region we similarly denote the supercharges by
\eqn\rtyp{~^{L_0}F^{J^3}_{\bar L_0}=^{\pm}F^{\pm}_{\pm \half}.}
These have eigenvalues $\pm \half$ under the action of
\eqn\szxc{\eqalign{                     L^{\rm far}_0&={2}\p_5,\cr
                      J_{\rm far}^3&=\p_\phi . }}
Employing the transformation to global coordinates for the $AdS_3$
one finds
\eqn\tti{ {P^3 \over 3a }H=\bar L^{\rm far}_0-\half(\bar
L^{\rm far}_1+\bar L^{\rm far}_{-1}).}
It then follows using \wsz\
that
\eqn\aprs{~^\pm Q^\pm=^\pm F^\pm_{\half}-^\pm F^\pm_{-\half}.}
Comparing \szxc\ and \szc\ with the help of the coordinate
transformations \srft\ one finds
\eqn\szc{\eqalign{ 2L^{\rm near}_0&=\bar L_0^{\rm far}-\half(\bar
L^{\rm far}_1+\bar L^{\rm far}_{-1}) +2 L_0^{\rm far}, \cr \bar
L^{\rm near}_0-L^{\rm near}_0&=  J^{3 \rm far}- L_0^{\rm far}, \cr
J_{\rm near}^3&= J_{\rm far}^3. }}

\newsec{Wrapped  branes}

In this section we describe some supersymmetric wrapped branes,
whose microstates will be argued in the next section to dominate the
entropy for large D0-charges.

\subsec{Probe supersymmetries}

Consider an M2 brane wrapped on the $S^2$ in the near geometry, and
sitting at the center $\rho=0$ of the $AdS_3$. As shown in \ssty\
this breaks half  the supersymmetry. Clearly this leaves unbroken
both $SU(2)$ rotations of the $S^2$ and  $U(1)$ rotations of the
$AdS_3$. It follows that the unbroken supercharges of the near
region must be
\eqn\rrpj{~^ +G^\pm_{\half}, ~~~~~~^-G^\pm_{-\half}   .}
Comparing with \prs\ we see that there are two globally unbroken
supercharges
\eqn\eso{ ~^+Q^+,~~~~^-Q^-.}
The existence of two unbroken global supercharges is a generic
feature of a supersymmetric probe configuration in backgrounds with
four supercharges.
%However when probe back reaction is included, the supergravity
%background is expected to adjust itself so that there are again four
%different unbroken supercharges, but with a different central
%charge.
Configurations preserving the same supersymmetries also exist which
carry angular momentum in the $AdS_3$ \ssty. These are given by
$\rho=\rho_0$, $x^5=x^5_0$, i.e.\ $\theta = -\tau + {x^5_0 \over
2}$.

\subsec{Explicit solutions}

A supersymmetric probe solution exists for M2 branes in the full
geometry. The Killing spinors of the metric \metric\ are given in
terms of the covariantly constant spinors of the $4d$
hyper-K$\ddot{a}$hler base space \GauntlettNW\
\eqn\hypermetric{ ds^2 = H^0 d\vec x^2 +
(H^0)^{-1}(dx^5+\omega^0)^2.}
The kappa-symmetry constraint for M2 branes static with respect to
the time $t$ requires the M2 brane to wrap a supersymmetric cycle in
the $4d$ base space, i.e. a cycle holomorphic for one of the complex
structures of the base space.

The geometry of the base space only depends on $H^0$ and is the same
for the full geometry and for the $AdS_3 \times S^2$ near geometry.
This is true even for the asymptotically flat geometry, as long the
constant terms in the harmonic functions satisfy $h^0=0$. The
natural choice for the moduli at infinity, i.e. the attractor moduli
for the $D4-D0$ black hole, satisfy this condition. A non-zero $h^0$
would require only a minor modification of this construction. Hence
the same equation $\rho = \rho_0$, $x^5 =x^5_0$ that defines a
supersymmetric $M2$ brane that wraps the sphere in $AdS_3 \times
S^2$ also defines a supersymmetric M2 brane in the full geometry.
This fact can be confirmed by looking at the complex structures of
the base space. A natural set of complex coordinates is given by
\eqn\complex{ U = e^{-i {x^5\over2}} \tanh {\rho\over 2}\,,\ \  V=
e^{i \phi} \sinh \rho \sin \eta.}
The M2 brane probe sits at $U=U_0$. Those are the only compact
holomorphic cycles in the geometry.

In the near geometry \raft, these probes move on a circle of
constant radius: $\rho=\rho_0$, $\theta = {x_0 \over 2} - \tau$, and
have angular momentum
\eqn\angm{
 J_\theta = -{1 \over 4\pi}{P^3 \over 3}\sinh^2{\rho \over 2}
}
Reduced to four dimensions, the M2 branes become ellipsoidal D2
branes, with the $D6$ and $\bar D6$ at the foci of the ellipsoid,
and worldvolume flux
\eqn\Dflux{ F=-{J_{\theta} \over 2} \sin\eta\, d\eta \wedge
(d\phi-2d\tau)-{1 \over 8\pi} {P^3 \over 3} \sin\eta \,d\eta \wedge
d\tau. }
The flux induces a $D0$ charge $q_0 = -J_{\theta}$, as expected from
the identification of charges between the near geometry and the
asymptotic geometry. This configuration is reminiscent of a
supertube \MateosQS, but unlike the supertube, here both the
fundamental string charge density and, proportional to this, the
four dimensional $J_\phi$ angular momentum density are zero; the
contributions from the Born-Infeld and Chern-Simons parts of the D2
Lagrangian cancel. Indeed, the M2 at $x_5 = x_5^0$ is transversal to
the M-theory circle, and does not carry any angular momentum $J_\psi
\, (=J_{\phi})$ along the $S^2$ it wraps. This is in contrast to the
D0 particle probes, which do carry $J_\phi$ angular momentum in the
$D6$-$\bar D6$ background. Finally, note that as in \gsyscqm\ these
ellipsoidal D2 branes move in the Calabi-Yau as particles in a
magnetic field proportional to $P$, because of the background
$F^{(4)}$ RR-field.

This analysis shows the existence of a new, larger moduli space of
multicenter configurations carrying $D4$-$D0$ charges: the moduli
space of a system formed by the $D6$-$\bar D6$ pair, a certain
number of $D0$ branes and several ellipsoidal $D2$ branes carrying
the extra $D0$ charge.

\newsec{Black hole entropy}

In this section we sketch how the Bekenstein-Hawking entropy emerges
from the counting of microstates for a system that includes the
ellipsoidal D2s.

The fastest way to understand the state-counting for this system is
to reduce it to a problem previously analyzed in \gsyscqm. In that
work, the superconformal chiral primary bound states of $D0$ branes
in the near-horizon $AdS_2\times S^2$ geometry of a $D4$-$D0$ black
hole were counted. A large degeneracy arises from non-abelian $D0$
configurations which puff up via the Myers effect and form a
$D2$-$D0$ bound state wrapping the horizon. This carries no $D2$
charge as measured at infinity. Such a $D2$-$D0$ configuration
couples to the background $F^{(4)}$ RR field and hence sees an
effective magnetic field on the Calabi-Yau. Hence the supersymmetric
ground states are highly degenerate lowest Landau levels.
Partitioning the $D0$ charge among such configurations produces an
exponential growth of states which exactly matches the
Bekenstein-Hawking entropy.

If we lift the calculation of \gsyscqm\ to M-theory, the near
horizon geometry becomes $AdS_3\times S^2$, the horizon-wrapped $D2$
becomes a horizon wrapped $M2$, and the $D0$ charge becomes orbital
angular momentum. Hence in M-theory the supersymmetric states of a
horizon wrapped $M2$ are highly degenerate lowest Landau levels on
the Calabi-Yau and reproduce the $D4$-$D0$ entropy.

In the current paper, we have deconstructed the $D4$-$D0$ black hole
into zero-entropy $D0$-$D6$-$\bar D6$ bits, which have a classical
moduli space.  It is natural to try to identify the black hole
microstates with the supersymmetric ground states of the moduli
space quantum mechanics (including the non-abelian interactions of
near-coincident $D0$s which are crucial for the Myers effect)). In
the $AdS_3\times S^2$ centered between the $D6$-$\bar D6$ pair, the
$D0$-branes have all been taken off to infinity. The entropy should
appear as the number of ways of reincorporating these $D0$s in this
$AdS_3\times S^2$. Using the wrapped $M2$ this counting proceeds
exactly as above and gives the desired result.

In addition to reproducing all the successes of \gsyscqm, the
current analysis potentially has several important advantages.  The
main observation of \gsyscqm\ was that the horizon-wrapped bound
states could account for the entropy. Reference \gsyscqm\  did not
attempt either to provide a systematic framework in which higher
order corrections to the entropy could be computed, or to derive or
justify the conjecture of \refs{\gt,\bsv} that states should be
counted in the global, rather than the Poincare,  time of the near
horizon $AdS_2$.   Perhaps these shortcomings can both be overcome
in the context of the deconstructed black hole quantum mechanics
discussed here.

\centerline{\bf Acknowledgements}

We would like to thank Miranda Cheng, Greg Moore and Toine Van
Proeyen for discussions. This work was supported in part by
DE-FGO2-91ER40654 and the Center for the Fundamental Laws of Nature
at Harvard University, and by the Belgian Federal Office for
Scientific, Technical and Cultural Affairs through the
``Interuniversity Attraction Poles Programme -- Belgian Science
Policy" P5/27 and by the European Community's Human Potential
Programme under contract MRTN-CT-2004-005104 ``Constituents,
fundamental forces and symmetries of the universe''. DVdB is
supported by an Aspirant scholarship from the FWO - Vlaanderen. XY
is supported by a Junior Fellowship from the Harvard Society of
Fellows.

 \listrefs
\end